\documentclass[sort&compress,onecolumn,3p,12pt]{elsarticle}

\usepackage{epsfig}
\usepackage{graphicx}
\usepackage{dcolumn}
\usepackage{amsmath}
\usepackage{enumerate}
\usepackage{pstricks}

\begin{document}

\begin{flushright}
IPPP /10/28 \\
DCPT /10/56\\
\end{flushright}
\vspace{2cm}
 
\title{High-$\gamma$ Beta Beams within the LAGUNA design study}

\author{Christopher Orme} 
%\ead{c.d.orme@durham.ac.uk}

\address{Institute for Particle Physics Phenomenology,\\ Department of Physics, Durham University,\\
  Durham, DH1 3LE, United Kingdom}

\begin{abstract}

Within the LAGUNA design study, seven candidate sites are being assessed for 
their feasibility to host a next-generation, very large neutrino observatory.
Such a detector will be expected to feature within a future European accelerator
neutrino programme (Superbeam or Beta Beam), and hence the distance from CERN is
of critical importance. In this article, the focus is a $^{18}$Ne and $^{6}$He 
Beta Beam sourced at CERN and directed towards a 50 kton Liquid Argon detector 
located at the LAGUNA sites: Slanic ($L=1570$ km) 
and Pyh\"{a}salmi ($L=2300$ km). To improve sensitivity to the neutrino mass ordering, these 
baselines are then combined with a
concurrent run with the same flux directed towards a large Water \v{C}erenkov 
detector located at Canfranc ($L=650$ km). This degeneracy breaking combination 
is shown to provide comparable physics reach to the conservative Magic Baseline 
Beta Beam proposals. For $^{18}$Ne ions boosted to $\gamma=570$ and $^{6}$He ions boosted 
to $\gamma=350$, the correct mass ordering can be determined at Slanic for all $\delta$ 
when $\sin^{2}2\theta_{13}>4\cdot 10^{-3}$ in this combination.

\end{abstract}
\maketitle

%%%%%%%%%%%%%%%%%%%%%%%%%%%%%%%%%%%%%%%%%%%%%%%%%%%%%%%%%%%%%%%%%%%%%%%%%%%%%%
%%%%%%%%%%%%%%%%%%%%%%%%%%%%%%%%%%%%%%%%%%%%%%%%%%%%%%%%%%%%%%%%%%%%%%%%%%%%%%

\newpage
\section{Introduction}

Results from a series of atmospheric~\cite{SKatm,atm}, solar~\cite{sol,SKsolar,SNO}, reactor~\cite{CHOOZ,PaloVerde,KamLAND} and long baseline 
accelerator~\cite{K2K,MINOS} 
neutrino experiments indicate that neutrinos are both massive and mix amongst themselves. A
combined analysis points to two approximate 2-neutrino mixing schemes, each parameterised by a mixing
angle and a mass-squared splitting. The extent to which these two schemes combine into a single
3-neutrino picture is controlled by the size of a third mixing angle, $\theta_{13}$. Defining
$\Delta m_{ji}^{2}=m_{j}^{2}-m_{i}^{2}$, a combined analysis~\cite{current} of all available data returns the best fit values
\begin{alignat}{4}
&\vert \Delta m_{31}\vert^{2} = 2.47 \times 10^{-3}\:\: {\rm eV}^{2} &\quad \mbox{and} \quad & \sin^{2}\theta_{23} =0.463 ~;&\notag\\
&\Delta m_{21}^{2} = 7.59 \times 10^{-5}\:\: {\rm eV}^{2} &\quad \mbox{and} \quad & \sin^{2}\theta_{12} =0.319~.& \notag 
\end{alignat}
The third mixing angle, $\theta_{13}$, is constrained to be
\begin{equation}
 \sin^{2} \theta_{13} < 0.016\:\:\: (0.053)\quad \mbox{at}\quad 1\sigma\:\:\: (3\sigma)~;
\end{equation}                                      
although some collaborations report a hint at low significance from a combined analysis of atmospheric, solar and
long-baseline reactor neutrino data~\cite{hint_Fogli,hint_Ge}.

The long term targets for the long baseline neutrino oscillation programme is to
determine the unknown neutrino mixing angle $\theta_{13}$; whether there is
CP-violation in the lepton sector, that manifests itself through a phase 
$\delta$; and to determine the sign of the atmospheric mass-squared splitting. 
The extraction of both the CP-phase and sign of the atmospheric mass-squared 
splitting require a three-neutrino analysis, the extent of the effect controlled
by $\theta_{13}$. Therefore sensitivity, optimisation and general experimental 
strategy are intrinsically related to the size of $\theta_{13}$. Running and near future experiments
will be the first to probe $\theta_{13}$ below the current experimental limit. If
$\theta_{13}$ is beyond the reach of these experiments, then intense sources of 
neutrinos from next generation Superbeams~\cite{Superbeam_US,T2K_up,rubbia_cern},
Neutrino Factories~\cite{nufact,nufactlow} or Beta Beams~\cite{zucchelli} will be necessary. 

A Beta Beam is a variation on the Neutrino Factory that instead sources neutrinos
from the decays of boosted radioactive ions. Proposed by Zucchelli in 
2001~\cite{zucchelli}, a Beta Beam distinguishes itself by sourcing beams of 
$\nu_{e}$ or $\bar{\nu}_{e}$ without contamination from other flavours and 
CP-conjugates. Through the analysis of the $\nu_{\mu}$ and 
$\bar{\nu}_{\mu}$ appearance channels, the Beta Beam provides a competitive 
physics reach on the unknown elements of neutrino mixing that have yet to be 
extracted from experiment. A Beta Beam will be a facility that will exploit existing, or proposed, ion 
production facilities and accelerator infrastructure. The major addition required
is a large decay ring to store the ions once they are boosted to the desired 
energy. In the laboratory frame, the neutrino flux is a function of both the 
ion decay Q-value and the Lorentz boost of the ions. Low Q-value ions; $^{18}$Ne 
and $^{6}$He; and high Q-value ions $^{8}$B and $^{8}$Li have been identified as 
the best candidate ions~\cite{Autin_et_al}. The focus of phenomenological study 
has been on a Beta Beam source at CERN using either the Super Proton 
Synchrotron~\cite{cernmemphys,Mauro,8fold,alternating,100_revisit} or an enhanced
1 TeV version, as required in some LHC upgrade 
scenarios~\cite{BB_upgrades,highergamma,singleion,BB_LENA,MB_4ions,MB_except,MB_Madrid,MB_INOpapers} 
(Other facilities could be the Tevatron at 
Fermilab~\cite{rubbia_FNAL,CPT_FNAL,FNAL_DUSEL}, or a re-fitted 
HERA-ring~\cite{DESY}.) In addition to specific setups, a number of optimisation
studies have also been carried out~\cite{lowtohigh,greenfield,minimal}.

In this article, the physics reach of Beta Beams directed towards Liquid 
Argon detectors in Europe will be simulated on the assumption that a 1 TeV 
machine will be available~\cite{LHC_upgrade}. The study of both low and high boost Beta Beams 
directed at large Water \v{C}erenkov (WC) detectors has been studied in detail and is
well 
understood~\cite{cernmemphys,Mauro,8fold,alternating,100_revisit,BB_upgrades,highergamma}. 
Water \v{C}erenkov detectors are best suited to short baselines since they only 
use the quasi-elastic events to reconstruct the signal. For longer baselines, and
hence higher energies, it is necessary to use and reconstruct the energy of 
multi-particle final state interactions, especially deep 
inelastic scattering events. For this reason, one must use active scintillator, 
calorimeters and projection chamber technologies for intermediate and very long
baselines. This study considers the GLACIER liquid argon detector~\cite{GLACIER}
as a far detector for the baselines  
CERN-Slanic ($L=1570$ km) and CERN-Pyh\"{a}salmi ($L=2300$ km) for 
$\gamma=350,350$ and $\gamma=570,350$ $^{18}$Ne and $^{6}$He Beta Beams. 
 (The CERN-Boulby baseline with a non-WC detector was discussed
in~\cite{singleion} for different exposures and so the very similar baseline for Sierozsowice
($L=950$ km) will not be included here.)These two 
baselines will then be combined with the same neutrino flux directed towards a 
large Water \v{C}erenkov detector based at Canfranc ($L=650$ km) to examine the possible 
broad physics reach.  

This work is part of the LAGUNA design study: an EC-funded project to assess the
feasibility of underground laboratories in Europe capable of housing a future,
large neutrino observatory. In addition to long baseline physics, the 
detectors aim to improve the bounds on proton decay half-lifes up to 
$\sim 10^{35}$ years to test a number of theoretical models; to detect neutrinos
from astrophysical objects and Cosmological sources; continued study of solar
and atmospheric neutrinos; and new sources such as from Dark Matter 
annihilations in the Earth's centre or Sun's core. Work towards understanding
detector response to various neutrino sources in currently on-going. This work
is one of a series of studies~\cite{BB_LENA,SB_LENA,rubbia_cern} examining the 
physics reach on long baselines
experiments with the LAGUNA network of detectors based on available 
information. These studies will help prioristise the potential host laboratories
for the future neutrino observatories.

The remainder of this article is organised as follows; in Sec.~\ref{S:baselines},
the baselines available within Europe as part of the LAGUNA design 
study~\cite{LAGUNA} will be listed and the basic phenomenological strategy for 
short and intermediate baselines will be summarised. In Sec.~\ref{S:single}
a description of the simulations carried out is made followed by the results for
Beta Beams directed along single baselines. In Sec.~\ref{S:twobase}, the 
simulations from Sec.~\ref{S:single} are augmented with an additional Beta Beam directed towards 
a large Water \v{C}erenkov based at Canfranc. The results are then discussed in relation to
Magic Baseline Beta Beams in Sec.~\ref{S:discussion}. Finally, in 
Sec.~\ref{S:summary}, the study is summarised.  

%%%%%%%%%%%%%%%%%%%%%%%%%%%%%%%%%%%%%%%%%%%%%%%%%%%%%%%%%%%%%%%%%%%%%%%%%%%%%%
%%%%%%%%%%%%%%%%%%%%%%%%%%%%%%%%%%%%%%%%%%%%%%%%%%%%%%%%%%%%%%%%%%%%%%%%%%%%%%

\section{Laguna sites hosting Beta Beam far detectors}\label{S:baselines}

Detector R\&D is not part of the LAGUNA Design Study~\cite{LAGUNA}; 
however the physics reach of these sites with respect to any European accelerator
neutrino programmes will be critical in any decision making process. Since 
future European Superbeams and Beta Beams will be one of the main users for 
these detector technologies; the distance from CERN will be a critical
factor in the final decision. However, since the size of $\theta_{13}$ and the 
number of future facilities and their scale is unknown, it is very hard
to determine what is the optimal baseline and detector choice. More precisely, 
optimisation of a facility can only be carried out once its purpose and 
circumstance is defined. Since $\theta_{13}$ is unknown, and (specifically) the role in which a
Beta Beam will take in a long term experimental strategy is uncertain; 
optimisation of the Beta Beam is not a well defined process. A typical strategy,
therefore is to aim for a broad physics reach using the ability to rule out
$\theta_{13}=0^{\circ}$, $\delta = 0^{\circ}$ and $180^{\circ}$, and determine the
correct neutrino mass ordering as indicators for a particular experimental 
setup; an approach also adopted here.  How one achieves this is dependent, amongst other things, on the 
choice of baseline (or baselines) and detectors technologies.

Within Europe there are three detector options being considered: a single volume 
Liquid Argon time projection chamber (LAr) known as GLACIER~\cite{GLACIER};
a 500 kton Water \v{C}erenkov (WC) known as MEMPHYS~\cite{Memphys} suitable only for energy reconstruction at short baselines where 
quasi-elastic events dominate;  and a 
non-segmented liquid scintillator detector known as LENA~\cite{LENA}. The 7 
laboratory sites being considered within LAGUNA to house these detectors are 
listed in Tab.~\ref{T:sites} along with their distances from CERN and 
corresponding first oscillation maximum energies. It clearly seen that a wide 
range of neutrino baselines are possible in Europe ranging from $L=130$ km to 
$L=2300$ km.

To develop a strategy, or interpret results of numerical simulations, for particular baselines
it is useful to consider an 
analytical approximation to the oscillation probability. With the definitions 
$\alpha \equiv \frac{\Delta m_{21}^{2}}{\Delta m_{31}^{2}}$ and  $\Delta 
\equiv \frac{\Delta m_{31}^{2}L}{4E}$; the
assumption of a constant matter density profile along the baseline 
$A\equiv \frac{2\sqrt{2}G_{F}n_{e}E}{\Delta m_{31}^{2}}$; and by expanding in all the small parameters one finds~\cite{probability}

%%%%
\begin{equation}
P(\nu_{e}\rightarrow\nu_{\mu}) \simeq \sin^{2}2\theta_{13}\cdot T_{1} +\alpha 
\cdot \sin\theta_{13}\cdot (T_{2}+T_{3}) +\alpha^{2}\cdot T_{4} ~,
\end{equation}
%%%%
where
%%%%
\begin{alignat}{1}
T_{1} =& \sin^{2}\theta_{23}\cdot \frac{\sin^{2}[(1-A)\cdot\Delta]}{(1-A)^{2}}~, 
\notag \\
T_{2} =& \sin\delta_{CP}\cdot \sin 2\theta_{23}\cdot \sin\Delta 
\frac{\sin(A\Delta)}{A}\cdot \frac{\sin((1-A)\Delta)}{(1-A)}~,\notag \\
T_{3} =& \cos\delta_{CP}\cdot \sin 2\theta_{23}\cdot \cos\Delta 
\frac{\sin(A\Delta)}{A}\cdot \frac{\sin((1-A)\Delta)}{(1-A)}~,\notag \\
T_{4} =& \cos^{2}\theta_{23}\cdot\sin^{2} 2\theta_{12}
\frac{\sin^{2}(A\Delta)}{A^{2}}~;
\end{alignat}
%%%%

%%%%%%%%
\begin{table}
\begin{center}
\scalebox{0.9}{%
\begin{tabular}{lcc}
\hline
\hline
{\bf Location} & {\bf Distance from} & {\bf Energy 1st Osc Max.} \\
     & {\bf CERN [km]} & {\bf [GeV]}\\
\hline
Fr\'{e}jus (France) & 130 & 0.26 \\
Canfranc (Spain) & 630 & 1.27 \\
Umbria (Italy) & 665 & 1.34 \\
Sierozsowice (Poland) & 950 & 1.92 \\
Boulby (UK) & 1050 & 2.12 \\
Slanic (Romania) & 1570 & 3.18 \\
Pyh\"{a}salmi (Finland)& 2300 & 4.65\\
\hline
\hline
\end{tabular}
}
\end{center}
\caption{Potential sites being studied with the LAGUNA design 
study~\cite{LAGUNA}. The Umbria site is of interest if an upgrade to the CNGS 
beamline is pursued.}\label{T:sites}
\end{table}

%%%%%%%
As is well known, and clear from this expression, the determination of $\theta_{13}$ and $\delta$ is severely affected by parameter correlations
and degeneracies~\cite{degeneracies}. For a given bin at fixed baseline, up to 8 possible parameter sets can fit the data. The challenge for a future 
long baseline experiment is to successfully resolve these degeneracies and push for a good physics return over the sought region of parameter 
space. With a 1 TeV machine, it is in principle possible to source a Beta Beam from very short long-baselines ($L=130$ km) up to very long long-baselines (Magic 
Baselines~\cite{MB_INOpapers}). A Beta Beam proposal therefore has access to two types of baseline in which some of the degeneracy can be naturally suppressed:
%%%%
\begin{enumerate}
%%%%
\item At short long-baselines ($L < 700$ km say), the matter effect is sufficiently small so that the true ($\theta_{13},\delta$) and the fake solution corresponding
to the incorrect mass ordering are close together. Consequently, the sensitivity to $\theta_{13}$ and $\delta$ is typically very good, especially with the
availability of large Water \v{C}erenkov detectors. Using the above expression, the CP-violation contribution to the probability is maximal for 
$L/E = 515/{\rm GeV}$. A Beta Beam flux with Lorentz boost $\gamma=350$, matches well the CERN-Canfranc baseline ($L=650$ km)~\cite{MB_4ions}. 
%%%%
\item At the `Magic Baseline', cancellations leave only the atmospheric contribution ($T_{1}$) 
to the appearance probability and thus there is no $\delta$ dependence. Numerically, this 
is found to be at $L=7250$ km~\cite{MB_original}. Use of the Magic Baseline in isolation will therefore return a reasonably clean measurement of $\theta_{13}$ and 
sign$(\Delta m_{31}^{2}$). The excellent sensitivity to these parameters, which is typical to proposals incorporating it, is because of 
the proximity to the matter resonance which, with the larger cross-sections at high energy, can compensate for the $L^{-2}$ dependence of the un-oscillated 
neutrino flux.
%%%%
\end{enumerate}
%%%
Both these options are a staple in long baseline proposals since the effective removal of one of the unknown parameters helps greatly with degeneracy resolution.
However, a extra baseline will be always be needed if one is to have a competitive reach on all three physics indicators: discovery of non-zero $\theta_{13}$,
CP non-conservation, and ability to rule out the incorrect neutrino mass ordering. For Beta Beams, the optimal choice for a second detector has not be determined 
rigorously, although several possibilities have been put forward for accompanyment of the Magic Baseline~\cite{MB_4ions,MB_except,MB_Madrid}. 

In this article, a Liquid Argon detector is considered at the two longest baselines within LAGUNA and therefore there is no natural suppression of the
degeneracy. By itself, a single baseline Beta Beam using a 50 kton detector will not return 
competitive sensitivities at the longer LAGUNA baselines owing to a suppression 
of the neutrino flux from the baseline ($ \propto L^{-2}$) and the presence of 
parameter degeneracies~\cite{degeneracies}. In isolation, binning the data to extract the oscillatory structure
of the appearance probability is the required strategy to break parameter degeneracies. The relative weight of the atmospheric ($T_{1}$), interference ($T_{2}$ and 
$T_{3}$)
and solar ($T_{4}$) contributions to the probability changes with the neutrino energy for a fixed baseline. Since different contributory features of 
the appearance probability dominate different bins and that the
location of parameter degeneracies are energy dependent, this strategy greatly aids degeneracy resolution. 

The above strategy was discussed in~\cite{singleion} for the CERN-Boulby baseline ($L=1050$ km) and is used in a number of other long baseline proposals, such as
the Wide Band Superbeam~\cite{WBB} and the low energy Neutrino Factory~\cite{nufactlow}. It was argued 
semi-analytically that combining data from first and 
second oscillation maximum, even for a neutrinos (or anti-neutrinos) only,
helps considerably in breaking the intrinsic energy degeneracy (or energy degeneracy~\cite{EC}). Degeneracy in $\delta$,
but not $\theta_{13}$ still remains, however. Incorporating data from surrounding
bins typically allows determination of the true $\delta$. Resolving the 
sign($\Delta m_{31}^{2}$) degeneracy at short to intermediate baselines is far 
harder. The difference in the probability for normal and inverted ordering
results from two effects: the $\sin\Delta$ in $T_{2}$ (which is present even in
the vacuum for 3-neutrino mixing); and from $(1-A)$ in $T_{1}$, $T_{2}$ and 
$T_{3}$. The discrepancy between normal and inverted orderings clearly increases
with baseline and neutrino energy. For the baselines considered here, the 
approach is therefore to extract the sign from high energy bins. Since these 
bins also contain information on $\theta_{13}$ and $\delta$, degeneracy will need
to be resolved. This, again, is achieved through the combination with the low
energy bins where the probability splitting between mass orderings is much 
smaller and hence the degenerate solution is much closer to the true solution. 
The limit on the sensitivity to the mass ordering is then limited by 
the available neutrino flux.

In the following section, the effectiveness of the above strategy will be 
explored for a Beta Beam directed along the CERN-Slanic and CERN-Pyh\"{a}salmi baselines. The event rates from the low energy bins are an important component of this approach, 
however, they are small owing the small detector mass and cross-sections. Therefore, following the initial analysis, the Liquid Argon baselines will be 
put in combination with a large Water \v{C}erenkov based at the Canfranc laboratory. As indicated above, this will provide a clean measurement of 
$\theta_{13}$ and $\delta$ which can then be combined with the high energy bins of the Liquid Argon detectors. The reach on the correct mass ordering is
expected to improve significantly as the inclusion of the extra (short) baseline is akin to enhancing the event rate in the low energy bins. The analysis carried
out will be discussed in more detail in the following sections.

%%%%%%%%%%%%%%%%%%%%%%%%%%%%%%%%%%%%%%%%%%%%%%%%%%%%%%%%%%%%%%%%%%%%%%%%%

\section{Single baseline study}\label{S:single}

The physics strategy in this study exploits $\nu_{e}$ and $\bar{\nu}_{e}$ beams 
sourced from the decays of boosted $^{18}$Ne and $^{6}$He. With production and
acceleration of the ions based at CERN, the physics will be simulated for a 
50 kton Liquid Argon detector located at the Slanic
($L=1570$ km) and Pyh\"{a}salmi ($L=2300$ km) mines. In the first instance only 
these baseline will be considered for a 5 year run of $\nu_{e}$ and 5 years
of $\bar{\nu}_{e}$ for the two boost pairings 
$(\gamma_{\nu},\gamma_{\bar{\nu}})=(350,350)$ and $(570,350)$. These boost assignments make the standard assumption that a 1 TeV Super Proton Synchrotron
will be available for a Beta Beam based at CERN~\cite{LHC_upgrade}. In the following 
section, the simulations will also include the combination with a larger Water 
\v{C}erenkov detector based at Canfranc ($L=650$ km) exposed to the same 
neutrino source.

The current R\&D in Liquid Argon detector technology is working towards a target
mass of 100 kton~\cite{GLACIER}. However, in this article a 50 kton detector is 
used for the following reasons
%%%
\begin{itemize}
\item To bring it in line with other non-WC based Beta Beam studies in the literature. 
\item The sought level of $^{18}$Ne decays along the straight section of the 
storage ring for the $\gamma=100,100$ proposal~\cite{Mauro} currently is beyond reach
by a factor 20, using ISOLDE techniques~\cite{Elena}. A 100 kton detector is assumed but with a factor of 2
used to offset a smaller than sought decay rate.
\end{itemize}
%%% 
The ion decay rates in the straight sections of the decay ring are currently 
unknown. R\&D towards production of $^{18}$Ne and $^{6}$He has focussed on the
use of ISOLDE techniques within the EURISOL design study~\cite{eurisol} 
with target rates of $1.1\times 10^{18}$ and $2.9\times 10^{18}$ yearly decays 
for a $\gamma=100,100$ machine for $^{18}$Ne and $^{6}$He respectively. 
Currently, the projected rates are a factor 20 short for $^{18}$Ne and about 2 
short for $^{6}$He~\cite{Elena}. This in part is because ISOL techniques, as applied by Nuclear Physicists, are optimized for more exotic nuclides that are
of little interest for Beta Beams~\cite{Don_Lin_Nufact}. The route to the sought decay rate for $^{18}$Ne could be through `direct production', for example
through the $^{16}$O($^{3}$He,n)$^{18}$Ne reaction~\cite{Direct}. 

Atmospheric neutrino events are skewed below 1 GeV and cause a problem for sourcing decays for the short baselines and need to suppressed with a 
restrictive duty factor in the decay ring ($\sim 10^{-3}$). However 
this constraint can be loosened for longer baselines where neutrino energies $E_{\nu}> 1$ GeV
are more important. This action will be necessary to reclaim the decay rate for high-$\gamma$ machines 
which, to first order, scales as $\gamma^{-1}$. (The impact of how much one inhibits the atmospheric neutrino background at the longer baselines has yet to 
studied in detail.) It may be possible to also claim some deficiency in the event rate through loosening the duty factor. 
To aid comparison with the latest
studies, in this article the assumption that $3\times 10^{18}$ useful ion decays
per straight section of the decay ring can be sourced per year for each ion as 
suggested in~\cite{Don_Lin_Nufact}, and used in~\cite{MB_4ions}, will be adopted.

For short and intermediate baselines, the aim is to exploit the energy 
dependence of the oscillatory structure of the appearance signal to help break 
the parameter degeneracies and push for a good physics reach. This is achieved 
through binning the signal to separate out the low and high energy appearance 
events so that the different strengths of the solar, interference and 
atmospheric contributions to the appearance probability can be observed. To this 
end, the detector has been assumed to possess a low energy threshold of 0.4 GeV 
and has binned up to 2.0 GeV in 0.2 GeV intervals. All bins thereafter are in 0.5 
GeV intervals up to the maximum laboratory energy of the neutrinos. At present, 
detector response for incident Beta Beam fluxes have not been simulated for the 
GLACIER Liquid Argon detector. Work in this respect in ongoing, with response 
data soon to be available for Superbeams directed at this 
technology~\cite{respon}. In its absence, a flat event efficiency of
80~\% has been taken for all channels with signal errors set at 2.5~\% and an 
energy resolution of $\sigma(E_{\nu}) = 0.15E_{\nu}$\footnote{These assignments are 
conservative with studies for FLARE~\cite{Flare} indicating that below 1 GeV, 
resolution could be as good as 2~\%, whilst above 2 GeV 
$\sigma(E_{\nu}) \simeq 0.10E_{\nu}$.}.
%%%%%%%%%%%
\begin{figure*}[t]
\begin{center}
\includegraphics[width=15cm]{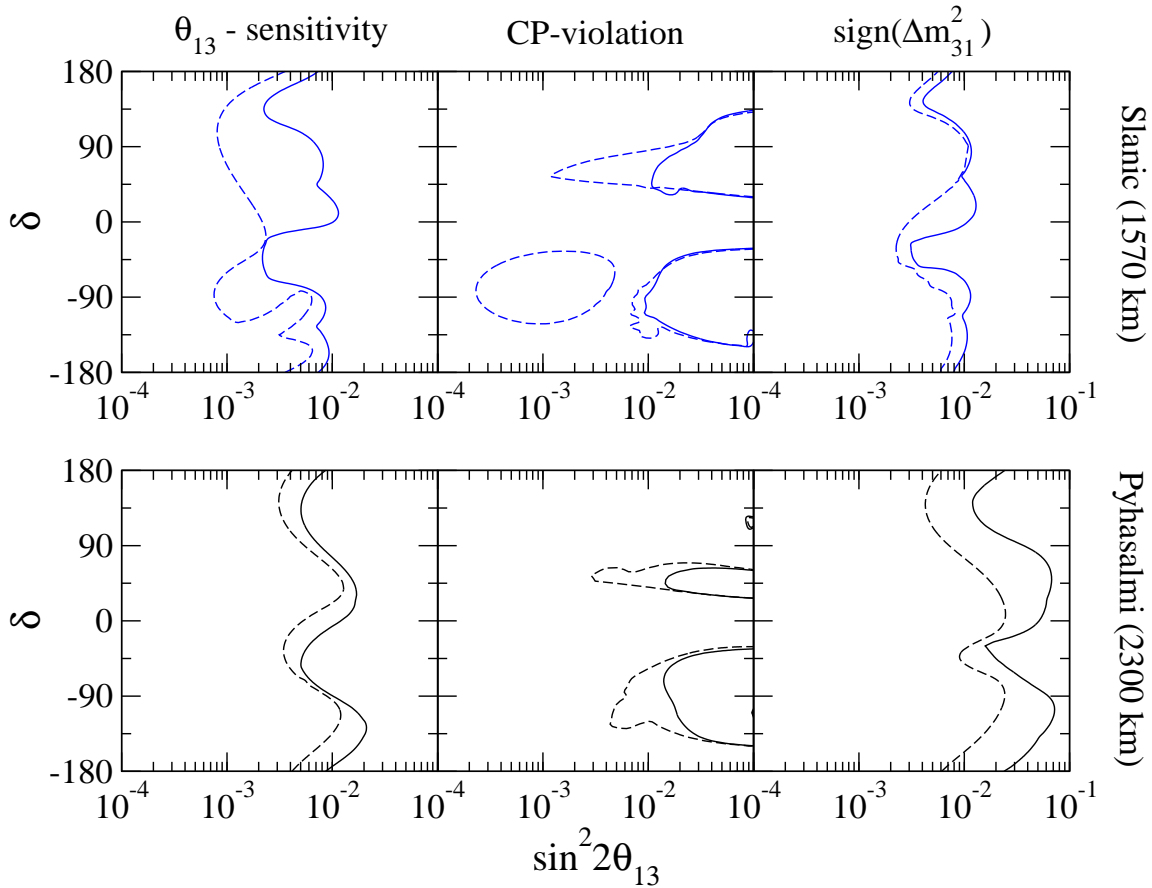}

\end{center}
\caption{$3\sigma$ C.L. contours for discovery for non-zero $\theta_{13}$ (left),
 CP-violation (centre), and sign($\Delta m_{31}^{2}$) determination (right). 
In each plot, the solid line corresponds to a $\gamma=350,350$ $^{18}$Ne and 
$^{6}$He Beta Beam for a 50 kton Liquid Argon detector located at 
Slanic ($L=1570$) (top line); and 
Pyh\"{a}salmi ($L=2300$ km) (bottom line). The dashed lines correspond to a 
$\gamma=570,350$ $^{18}$Ne and $^{6}$He Beta Beam.}\label{Fi:single}
\end{figure*}

All simulations in this study have been carried using the GLoBES long baseline 
simulation software~\cite{globes}. The known oscillation parameters have been
fixed to their current central values, taken from~\cite{current}, and are always
marginalised over. The exception is $\theta_{23}$ which is fixed to $45^{\circ}$ so that the octant degeneracy is absent.
 The errors have been set to 1~\% for the solar parameters and
5~\% for the atmospheric parameters and matter potential. Negligible background 
from neutral current events is expected and has been set to 0.1~\%. 
The 1 degree of freedom convention is adopted for all sensitivity plots in this 
paper. 

\subsection{Results}

The physics reach of the two baselines considered in summarised is 
Fig.~\ref{Fi:single}. The 3$\sigma$ confidence level contours are shown for 
sensitivity to non-zero $\theta_{13}$ (left), CP non-conservation (middle), and to resolving the 
sign($\Delta m_{31}^{2}$) ambiguity (right). The analysis takes into account the impact 
of intrinsic and the sign($\Delta m_{31}^{2}$) degeneracies.  The top line of Fig.~\ref{Fi:single} shows the results for the 
CERN-Slanic baseline, whilst the bottom display the outcome of the CERN-Pyh\"{a}salmi simulations.

The best sensitivity to non-zero $\theta_{13}$ and CP-violation is found for the CERN-Slanic
baseline. This is to be expected since, with the same source, the flux is larger for this baseline. The weaker matter
effect means that the sign($\Delta m_{31}^{2}$) degeneracy interferes less with these measurments.
For the $\gamma=570,350$ boost pair, non-zero $\theta_{13}$ can be seen down to $\sin^{2}2\theta\sim 10^{-3}$, and, for both boost pairs, at all values of $\delta$
for $\sin^{2}2\theta_{13}>10^{-2}$. However, there is a marked difference between the two boost pairings for sensitivity to CP-violation. For the
$\gamma=350,350$ pair, the ability to rule out $\delta =0^{\circ}$ or $180^{\circ}$ 
is restricted to $\sin^{2}2\theta_{13}>10^{-2}$, but increasing the boost of the $^{18}$Ne ions
returns a large region of parameter space for $\delta <0^{\circ}$ and centred on $\sin^{2}2\theta_{13} =10^{-3}$. There is little enhancement on the region for 
low boost paring. This is suggestive that degeneracy is a problem for the $\gamma=350,350$ boost pairing; especially 
for $\sin^{2}2\theta_{13}\sim 5\cdot10^{-3} - 1\cdot 10^{-2}$ where there is a gap in CP-violation sensitivity.  
The lower event rates for the longer 
baselines mean that the data is insufficient to reduce the significance of some 
degenerate 
solutions. Although, the ability to rule out the incorrect mass ordering is poor relative to Neutrino Factories, for a Beta Beam it is not intrinsically bad.
For the high boost run, the correct ordering can be indentified down to $\sin^{2}2\theta_{13}=2\cdot 10^{-3}$, with determination for all values of $\delta$
for $\sin^{2}2\theta_{13}>10^{-2}$ in both cases. Although the increase in the $^{18}$Ne boost improves the reach, it does not do so significantly. Increasing 
the boost makes data from higher enegies available without significantly altering the event rate and composition at lower energies. Since
European baselines make use of these bins in combination, improving one without the improving other need not, and has not, dramatically 
improved the physics return. The low event rate at low energies is still insufficient to break the degeneracy for small values of $\theta_{13}$.

The physics return for the CERN-Pyh\"{a}salmi baseline is weaker for each of the physics indicators, with little sensitivity for $\sin^{2}2\theta_{13}<10^{-2}$.
Principally, this is due to the $L^{-2}$ dependence of the un-oscillated neutrino flux. In particular, the ability to determine the correct mass ordering is
very poor even given the large matter effect at this baseline. The true and incorrect mass ordering solutions will be sufficiently separated in $(\theta_{13},\delta)$
space; however, the low
event rate means that the solution regions at $3\sigma$ will be large and possibly merged together. When combined with a large solution region from the 
low energy bins (also owing to low event rates), the data is insufficient to break the degeneracy for small $\theta_{13}$. 

In summary, neither of the baselines
considered here have, in isolation, the right combination for degeneracy breaking ability and sufficient un-oscillated event rate to return a competitive 
physics reach on all physics indicators. One solution to this problem is to combine these
baselines with another beam. It has been proposed in~\cite{CPT_FNAL,Schwetz_small} to use combinations of Beta Beams and Superbeams at these shorter baselines
to break degeneracy and, in particular, improve the ability to determine the mass ordering. This strategy exploits the different forms of the appearance 
probability for the $\nu_{e}\rightarrow \nu_{\mu}$, $\bar{\nu}_{e}\rightarrow \bar{\nu}_{\mu}$,$\nu_{\mu}\rightarrow \nu_{e}$ and 
$\bar{\nu}_{\mu}\rightarrow \bar{\nu}_{e}$ channels which in turn affects the location of the degenerate solutions for different oscillation channel pairs. 

In the next
section, a variant on this approach will be adopted. Specifically, a second Beta Beam baseline will be introduced in view of returning 
sensitivity to $\theta_{13}$ and $\delta$ to much greater precision. This is extrapolated from the strategy advocated earlier of using low and high energy bins
and is motivated from the results for the Slanic baseline indicating that improvement in the high energy neutrino event rate is insufficient to significantly 
improve the hierarchy reach. Increasing the ion boost does not significantly improve the event rate at low energies; however, introducing a second Beta Beam
at a shorter baseline, and with a large Water \v{C}erenkov will have the desired effect. This shorter baseline can be used to locate
the true solution accurately to the level $\sin^{2}2\theta_{13}\sim 10^{-4}$, effectively replacing the low energy bins of the (now) far detector.
With this information the longer baseline can then 
rule out the incorrect solution from the sign($\Delta m_{31}^{2}$) degeneracy for much smaller $\sin^{2}\theta_{13}$. This idea will be presented in the 
next section.

%%%%%%%%%%%%%%%%%%%%%%%%%%%%%%%%%%%%%%%%%%%%%%%%%%%%%%%%%%%%%%%%%%%%%%%%

\section{In combination with a large Water \v{C}erenkov}\label{S:twobase}

As discussed in Sec.~\ref{S:baselines},
for a single baseline, the strategy is to use the complementary information from 
low and high energies to break degeneracies and push to physics reach at small 
$\theta_{13}$. However, for the intermediate baselines considered here, owing to 
smaller detector sizes and increasing 
sign($\Delta m_{31}^{2}$)-degeneracy dominance, the $\theta_{13}$ and $\delta$ 
reaches are relatively poor compared to the Water \v{C}erenkov Beta Beams. This was demonstrated through the simulations of the previous section which indicated
that degeneracy persisted for the intermediate baselines of Europe. A straightforward way to overcome this is to increase the exposure so that degenerate solutions
lose statistical significance. However, for the Beta Beam, demands on production and ion storage impose restrictions on scaling up of this kind.  

The challenge is improve the event rate at the low energies and then use this data in combination with high energy neutrinos. This approach is distinct from
the standard technique to resolve the neutrino mass ordering. It is typical to consider very long baselines where the matter effect is strongest and the closeness
to the matter resonance allows the recovery of flux reduced by its $L^{-2}$ dependance. Most interest centres on the Magic Baseline where CP-violation effects
vanish from the appearance probability, leaving only the atmospheric contributions. There have been a number of studies using a second Beta Beam in addition to a 
Magic Baseline Beta Beam to help break degeneracies. Thus far, a Magic 
Baseline/short baseline combination has been studied in~\cite{MB_except,MB_4ions} and a 
Magic Baseline/long baseline combination in~\cite{MB_Madrid}. The strategy employed in this section is to deliberately place the far detector at a baseline
where the degeneracy is strong. In a short baseline/intermediate baseline configuration the short baseline will provide the sensitivity to $\theta_{13}$ and 
$\delta$ down to the $\sin^{2}2\theta_{13}\sim 10^{-4}$ level. This information 
can then used to rule out degenerate solution regions in ($\theta_{13},\delta$) 
space for the longer baseline. (For a short baseline, the 
sign($\Delta m_{31}^{2}$)-degenerate solution will be very close to the true 
solution. As the matter increases, the two solutions diverge.) The combination of
data sets is expected to improve the sensitivity considerably. This is demonstrated in the following simulations.

\subsection{Results}

%%%%%%%%%%%%%%
\begin{figure*}[t]
\begin{center}
\includegraphics[width=15cm]{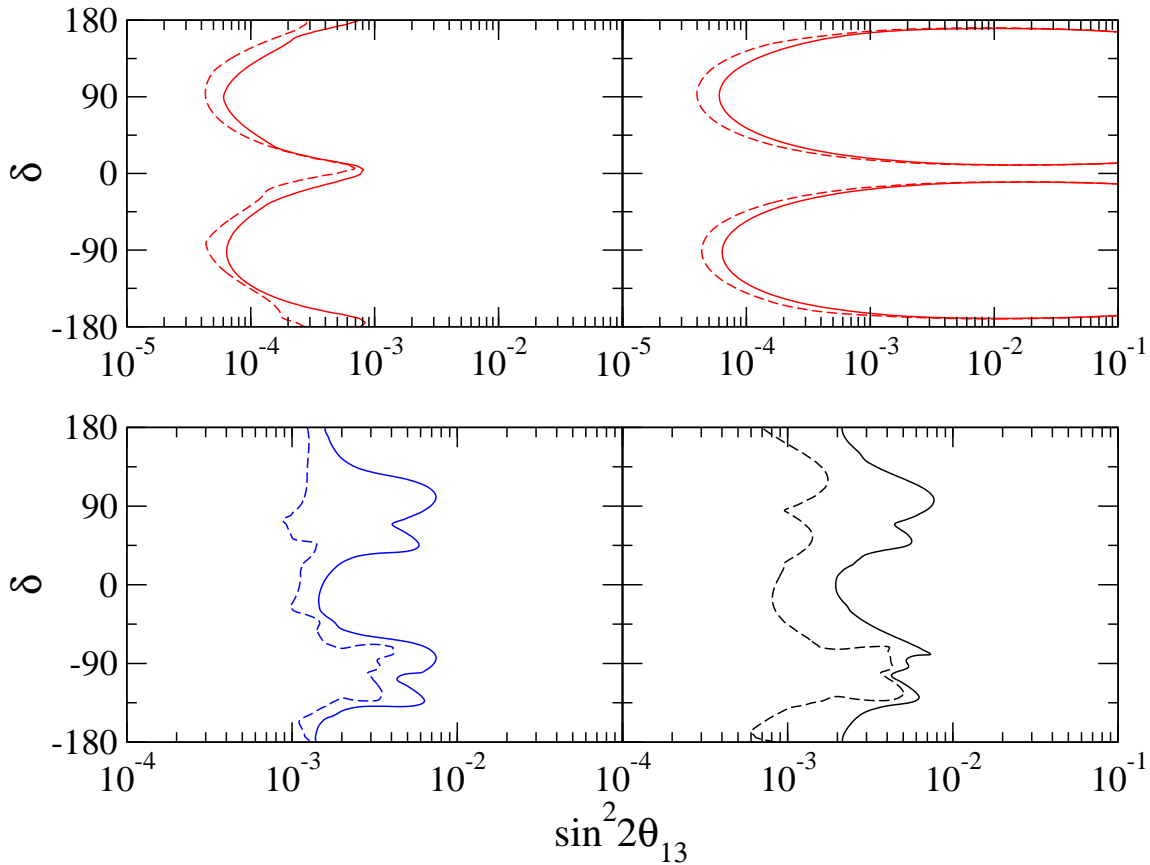}

\end{center}
\caption{$3\sigma$ C.L. contours for discovery (for the far detector at Slanic) for non-zero $\theta_{13}$ (top left) and
 CP-violation (top right). On the bottom line $3\sigma$ C.L. contours for sign($\Delta m_{31}^{2}$) determination for Slanic (left)
and Pyh\"{a}salmi (right) 
In each plot, the solid line corresponds to a $\gamma=350,350$ $^{18}$Ne and 
$^{6}$He Beta Beam and the 
dashed lines correspond to a $\gamma=570,350$ $^{18}$Ne and $^{6}$He Beta Beam.}
\label{Fi:dual}
\end{figure*}
%%%%%%%%%%%%%%

In Fig.~\ref{Fi:dual}, the $3\sigma$ confidence levels for the three physics 
indicators are shown for a large Water \v{C}erenkov, based at Canfranc,
in combination with the same Liquid Argon detector as previous for the the 
Slanic and Pyh\"{a}salmi laboratories. The Water \v{C}erenkov
dominates the $\theta_{13}$ and $\delta$ sensitivities and the reaches are 
essentially identical in both combinations. Therefore only the results for Slanic are shown here. 
The CP-discovery plot is both smooth
and symmetric. There is no residue intrinsic degeneracy at $\sin^{2}2\theta_{13}\sim 10^{-2}$ as its location is different for each baseline and therefore
can be resolved. For both baseline pairs, non-zero 
$\theta_{13}$ and $\delta$ distinguishable from $0^{\circ}$ and $180^{\circ}$ can be achieved down to $\sin^{2}2\theta_{13}=5\cdot 10^{-5}$. 

The important result is that the combination of baselines does indeed improve 
the ability to rule out the incorrect mass ordering beyond the capability of either
baseline separately. Specifically, for $\gamma=350,350$ and for both 
combinations, the correct mass ordering can be extracted for all $\delta$ down 
to $\sin^{2}2\theta_{13}\sim 6-7 \cdot 10^{-3}$. For $\gamma=570,350$ this 
improves to $\sin^{2}2\theta_{13}\sim 4 \cdot 10^{-3}$ which is the level 
reported for the conservative Magic Baseline proposal in~\cite{MB_4ions}. Even 
for $\gamma=350,350$, there is substantial resolving power at the 
$4\cdot 10^{-3}$ level indicating that a more more minimal Beta Beam is capable
of returning similar physics.

%%%%%%%%%%%%%%%%%

\section{Discussion}\label{S:discussion}

The simulations presented in this article indicate that a two-baseline Beta Beam
configuration using a short and intermediate baseline can obtain a similar
physics reach as conservative Magic Baseline proposals, such as 
in~\cite{MB_4ions}.  Long baseline experiments incorporating the Magic Baseline aim to exploit 
the absence of the CP-violation at this baseline along with the nearby resonance 
associated with atmospheric neutrino mixing. Since such an environment provides
a relatively clean measurement of the mass ordering, usually to very small $\theta_{13}$, 
Beta Beam studies aiming to achieve the best reach in this respect typically
exploit it~\cite{MB_4ions, MB_except,MB_Madrid,MB_INOpapers}. It is instructive 
to highlight the reason why a more minimal setup can return similar sensitivity
to the conservative Magic Baseline proposals.

The crucial point to realise is that the sought features of the baseline, 
namely no CP-phase and proximity to the matter resonance, are features of the 
appearance probability. In an experiment, one measures the event rate which is a
convolution of the unoscillated neutrino flux at detector, a cross-section, 
detector efficiencies and the probability. Having a signal clean from CP-violation
is of little use if the event rate is too low to provide a competitive 
sensitivity. The unoscillated flux has a $L^{-2}$ dependence which greatly 
reduces the flux at very long baseline. This reduction can be recovered through 
the larger cross-sections at high energy and the matter enhancement of the 
appearance probability. On the assumption that the same number of ion decays is 
available for the very long baselines as for the short, the number of appearance
events is roughly independent of the baseline at very long 
baselines~\cite{MB_INOpapers}. In this instance, the ability to extract the mass
ordering will peak where the signal is cleanest from $\delta$ `contamination': the Magic Baseline. However, a 
second baseline is always necessary to have access to the CP-violation. In which
case, it may be possible to use synergy between baselines to extract the sign of
the mass splitting at a similar level as demonstrated in this study. 

Never-the-less, the sensitivity presented here can still be beaten by most Magic 
Baseline Beta Beams considered.  The point raised in~\cite{MB_4ions} is that 
construction restrictions on the size of the decay ring could result in
a considerable reduction of the flux at very long baseline in a realistic 
proposal. (Longer baselines require the decay ring to dip at larger angles to 
the surface so that the maximum depth of the ring could be beyond 2km if very powerful superconducting magnets are not available.) 
If one imposes a limit on the maximum depth of the decay ring, then there is a maximum 
baseline for which one can source $3\times 10^{18}$ ion decays per year in the straight 
section. For all baselines longer than this, the length of the straight sections
need to be curtailed. If the flux at the Magic Baseline is sufficiently reduced,
there will become a point when the $\delta$ cleanliness of the signal is insufficient to 
achieve the best physics reach, even for a single baseline. At such a point, 
combinations involving synergy would then be sought to obtain a broad overall 
physics reach. By choosing to use $^{18}$Ne and $^{6}$He only, this study was 
able to expose both detectors simultaneously to the neutrino flux. In studies 
using all four candidate ions, one effectively cuts the event rate in half since
the low-Q and high-Q pairs need to be run separately and irradiate only the 
appropriate detector. The combination of the synergy between baselines and the 
higher event rate from using only two ions is the reason why the relatively 
minimal configurations discussed here have returned similar physics to  
conservative Magic Baseline proposals.
  
%%%%%%%%%%%%%%%%%%%%%%%%%%%%%%%%%%%%%%%%%%%%%%%%%%%%%%%%%%%%%%%%%%%%%%%

\section{Summary}\label{S:summary}

Assessing the physics reach of long baselines with far detectors based at the 
potential LAGUNA sites is of critical importance for strategic decisions towards
accelerator and non-accelerator neutrino physics in Europe. Recently, as part of 
the LAGUNA design study, work studying the physics return 
for a Beta Beam with LENA~\cite{BB_LENA} as a far detector, and a high powered
Superbeam directed towards GLACIER~\cite{rubbia_cern} have been performed. In 
this article, this has been continued to include Beta Beam physics with 
GLACIER, but with the mass effectively reduced to 50 kton to 
offset any shortfall in the useful ion decay rate. R\&D towards the detector 
response for the incident beam is currently unavailable, and so the approach was
to be conservative in energy resolution and efficiency assumptions. 
Configuring the detector with a low energy threshold and binned in a manner to
extract the oscillatory structure of the signal, the physics for single and 
double baselines was simulated.  

The physics reach for a Beta Beam directred towards the Slanic Mine ($L=1570$ km)
and Pyh\"{a}salmi ($L=2300$ km) were simulated initially. The shorter baseline performed
best owing to its larger event rate and weaker sign($\Delta m_{31}^{2}$) degeneracy.
Non-zero $\theta_{13}$ could be established down to $\sin^{2}2\theta_{13}\sim 10^{-3}$ for the 
$\gamma=570,350$ facility. The effect of altering the boost of the $^{18}$Ne ion is large
for the Slanic baseline, especially for CP-violation. This indicates that degenerate solutions pose
a problem for the less energetic neutrinos. The effect was much less apparent for identifying the
correct mass ordering. This is because the increasing the boost gives access to higher energy neutrinos
without substantially increasing the flux at low energies. For Pyh\"{a}salmi, the sensitivities were much
weaker for all indicators with little return below $\sin^{2}2\theta_{13}=10^{-2}$. Although the matter effect
is larger at Pyh\"{a}salmi, the reduced flux from being more distant from source allows the degenerate
solutions to remain statistically significant for small $\theta_{13}$. 

These results indicated that the data from the Slanic and Pyh\"{a}salmi Liquid 
Argon detectors should be  combined with a concurrent run directed towards a large 
Water \v{C}erenkov based at Canfranc. Such an addition provides an almost clean 
measurement of $\theta_{13}$ and $\delta$ down to 
$\sin^{2}2\theta_{13}\sim 10^{-4}$. This information is then available for the 
longer baseline to remove the degenerate solutions and extract the mass ordering
from its data set. The reach on the neutrino mass ordering is competitive with 
the conservative two baseline configurations incorporating the Magic Baseline.   
The access to $\theta_{13}$ and CP-violation was dominated by the short baseline
with reach down to $\sin^{2}2\theta_{13} =5\cdot 10^{-5}$. 

In conclusion, the results of this study indicate that European Beta Beams can be provide competitive physics reach to move extensive proposals. 
The optimal Beta Beam configuration, therefore, has yet to be determined and should be investigated further, taking into account available knowledge on 
technological restrictions. More generally, these results demonstrate that the Magic Baseline is not mandatory for determining the sign of $\Delta m_{31}^{2}$: 
the combination of data from a short baseline with data heavy in degeneracy can be an equally as powerful phenomenological tool.

\section*{Acknowledgments}

This work was supported by the European Commission Framework Programme 7 Design 
study: LAGUNA, project number 212343. The EC is not liable for any use that 
maybe made of the information contained herein.

\end{document}